\documentclass[submission,copyright,colorlinks=true]{eptcs}
\usepackage[T1]{fontenc}
\providecommand{\event}{FIT 2010} 
\usepackage{breakurl}             
\usepackage{amsmath,amsthm,amssymb,subfig}
\usepackage{microtype}
\usepackage[pdftex]{graphicx}
\newcounter{assumption}
\newcommand{\Assume}[1]{\smallskip\addtocounter{assumption}{1}%
	\noindent\framebox{\textbf{A\arabic{assumption}}}\,%
	\expandafter\newcommand\csname ref#1\endcsname%
	{\arabic{assumption}}}

\newtheorem{theorem}{Theorem}

\title{A Few Considerations on Structural and Logical\break Composition in Specification Theories}
\author{Axel Legay
\institute{INRIA Rennes, France}
\email{axel.legay@irisa.fr}
\and
Andrzej W\k{a}sowski
\institute{IT University of Copenhagen, Denmark}
\email{wasowski@itu.dk}
}
\def\titlerunning{A Few Considerations on Structural and Logical Composition\dots}
\def\authorrunning{Legay, W\k{a}sowski}
\begin{document}
\maketitle

\newcommand{\SPEC}{\textbf{S}}
\newcommand{\UNI}{\ensuremath{\mathsf U}}
\newcommand{\NULL}{\ensuremath{\mathsf \emptyset}}

\begin{abstract} Over the last 20 years a large number of automata-based
	specification theories have been proposed for modeling of discrete,
	real-time and probabilistic systems.  We have observed a lot of shared
	algebraic structure between these formalisms.  In this short abstract, we
	collect results of our work in progress on describing and
	systematizing the algebraic assumptions in specification theories. 
\end{abstract}

\section{Introduction}

Specification formalisms commonly support two main ways of combining
specifications: conjunction and parallel composition.  The former,
conjunction, is more common in specification logics (such as temporal logics
\cite{baier.katoen:2008}).  It focuses on combining distinct views on the same
system, or a component.  The latter, parallel composition, more often
discussed in process algebraic approaches \cite{handbook},  talks about structurally composing
two or more communicating systems.  

Recently a few specification theories have been
developed that support both parallel composition and conjunction (for example
\cite{DBLP:conf/hybrid/DavidLLNW10,caillaud.ea:2009:qest}).
Interestingly, placing both parallel composition and conjunction in the same theoretical framework raises
a lot of natural questions.  When presenting the results of 
\cite{DBLP:conf/hybrid/DavidLLNW10} and
\cite{caillaud.ea:2009:qest}, we have received a wide range of opinions about
the two operators, ranging from \emph{composition and conjunction are radically
different}, through \emph{they are easy to confuse, and hard to distinguish}
to extreme ones such as  \emph{conjunction and parallel composition are essentially the same}.
A more diligent insight into the properties of the two operators shows that
each of the opinions is in a way justified.  For example conjunction satisfies
the most essential axioms of parallel composition in some cases, and both
operators are computed as pruning results of a certain kind of product.
	
In this note we collect most important observations about
similarities, relations and differences between conjunction and parallel
composition in specification theories, as we have experienced in our prior
work.  
First, we show that \emph{parallel composition refines conjunction} for a
broad class of single-player specification theories.  Under the rather general
assumptions that conjunction is the greatest-lower bound with respect to
refinement, and the refinement is a precongruence for parallel composition, we
show that existence of a certain kind of universal specification suffices for
parallel composition to be always a stronger operator  than conjunction
(stronger as defined by the refinement relation).
This fact is easily observed using basic properties
of the order theory.  It is not necessary to assume any specific properties
about the specification language.  An interesting side-effect is that parallel
composition behaves much like conjunction under these assumptions.  For example
composing more systems to an existing assembly strengthens the specification
(in the sense that it decreases the set of models).

These observations apply out of the box to theories of Constraint Markov
Chains~\cite{caillaud.ea:2009:qest} and modal transition systems (scattered
across multiple
papers~\cite{Larsen89,DBLP:conf/esop/LarsenNW07,RBBCLP09,RG-sefm09}).  It
would also apply to a natural theory that one could build with finite automata
and language inclusion as refinement.  

The existence of a universal specification  requires an implicit closed-world
assumption of a kind, which tends to hold in specification
theories with a single-player semantics, like the above listed.  Such theories
do not distinguish between choices made by the component, and
choices set by the environment in which it operates.   In contrast
one \emph{cannot} always assume existence of a single largest specification when 
modeling under the open-world assumption.  
Consequently the above relation
between parallel composition and conjunction does not hold for two-player
specification theories such as Interface Automata \cite{dAH01}, Timed
Interfaces \cite{DBLP:conf/emsoft/AlfaroHS02} or Timed I/O Automata
\cite{DBLP:conf/hybrid/DavidLLNW10}. 
For these models we can spot an interesting duality between the two operators.

Type theories, and older specification theories used in verification typically
focus on \emph{universal} correctness using a \emph{pessimistic} composition, i.e.
two systems are compatible, if under no circumstances their execution can lead
to an error. In \cite{dAH01} Alfaro and Henzinger propose an optimistic
composition operator, which stipulates that two components are compatible as
long as there exists a context in which they will not fail. 
In \cite{DBLP:conf/hybrid/DavidLLNW10,DBLP:conf/emsoft/AlfaroH01} this duality
between 
optimistic--pessimistic is investigated further. In
\cite{DBLP:conf/hybrid/DavidLLNW10}
we observe that the pessimistic
composition is obtained by synthesizing a winning strategy for the player
representing the system, while the optimistic composition is obtained by
synthesizing a winning strategy for the player representing the environment.

Furthermore, the two compositions suit well two different use cases.  When an
implementer receives a contract specification for the component, the
environment assumptions are given, while some choices in the implementation
still remain under his control.  Thus if constructing specifications for
implementers the pessimistic composition is more suitable.   In practice this
means that conjunction should be computed as the strategy for the component,
since conjunction is a natural way of constructing a contract.

Dually, the user of the component typically cannot influence the component
itself, but can change the details of how it is used.  This is why the
optimistic composition is more suitable when synthesizing implementations for
the users of black-box components.  In this scenario parallel composition is
the natural composition operator, and since the usage scenario is the only
element under control, the composition is computed as the winning strategy for
the environment player. 


\section{A Single Player Setting}\label{sec:one-player}

In the following let \SPEC\ be a universe of specifications.

\Assume{c-refinement} Let $(\leq) \subseteq (\SPEC\times \SPEC)$ be a binary relation on this
universe.  We call it a \emph{refinement}.  Our refinement induces an
equivalence on specifications: $A\equiv B$ iff $A\leq B$ and $B\leq A$.

\Assume{c-preorder} We assume that $\leq$ is a pre-order (reflexive and
transitive).

\Assume{c-universal} We postulate
existence of a \emph{universal} specification \UNI, such that $A\leq \UNI$ for
all $A\in\SPEC$.  

\bigskip

\Assume{cConjunction} Let conjunction be a binary operator: $\land : \SPEC \times \SPEC \rightarrow
\SPEC$.

\Assume{c-conj-total} Conjunction is total : $A \land B$ is defined for all $A,B\in\SPEC$.

\Assume{c-conj-commutative} Conjunction is commutative: $A \land B \leq B \land A$ for all $A,B\in\SPEC$.

\smallskip\noindent Conjunction is the greatest lower bound with respect to $\leq$:

\Assume{c-conj-glb-1} $A \land B \leq A$ and $A \land B \leq B$ for all $A,B\in\SPEC$.

\Assume{c-conj-glb-2} If $C \leq A \text{ and }  C \leq B \text{ then }C \leq
	A \land B$

\bigskip
	
\Assume{c-par} Another operator $(\mid) : \SPEC \times
\SPEC \rightarrow \SPEC$ is called parallel composition.  

\Assume{c-par-total} Total: $A\mid B$ is defined for all $A,B
\in\SPEC$.\footnote{Totality of parallel composition is not strictly needed, but added to this note to simplify
assumptions of further theorems and properties --- in principle it is enough
to always assume existance of parallel compositions in question.}

\Assume{c-par-commutative} Commutative: $A\mid B \leq B\mid A$ for all $A$, $B$.

\Assume{cPrecongruence}
Refinement is a precongruence for the contexts defined by parallel composition, so:
\begin{equation}
	A \leq B \text{ implies that } A \mid C \leq B \mid
	C
\end{equation}

\smallskip\noindent We additionally require that \UNI\ is a unit of parallel composition:

\Assume{cUniversalParallel} $A \mid \UNI \leq A$ for all specifications $A$ 

\smallskip

At first the above may appear a
strong requirement, but in fact this is very natural, and holds in many
specification theories.  For example if '$\mid$' is a product of automata then
the product of $A$ with a universal automaton \UNI\ gives $A$
again.
In a more
complicated scenario of CCS-like synchronizations where alphabets of $A$ and
\UNI\ may differ, the refinement enforces alphabet equalization, which
normally is defined by $A \mid \UNI$ and automatically gives the above.

\begin{theorem}\label{thm:main}
	$A \mid B \leq A \land B$
\end{theorem}
\begin{proof} Note that $A\leq \UNI$ so by precongruence $A \mid B \leq \UNI
	\mid B \leq B$ (the latter by assumptions A11 and A13).  We conclude similarly that
	$A \mid B \leq A$ and since conjunction is the greatest lower bound $A
	\mid B \leq A \land B$.
\end{proof}


As an example, observe that assumptions of Thm.~\ref{thm:main} hold for a theory build around Constraint Markov
Chains \cite{caillaud.ea:2009:qest}, Modal Specifications (not fully built in any paper, but
results are scattered across more papers 
\cite{Larsen89,DBLP:conf/esop/LarsenNW07,RBBCLP09,RG-sefm09}), or
just finite automata with language inclusion as refinement. 

The condition A13, that $A \mid \UNI \leq A$, is in fact the sufficient and
necessary condition for Thm.~\ref{thm:main} to hold, in the following sense.

\begin{theorem}
If the universal element exists, and the refinement is a precongruence, and
conjunction is the greatest lower bound, but $A \mid \UNI \not\leq A$ 
then not necessarily $A\mid B \leq A\land B$.
hold.\end{theorem}

This is easy to show by contradiction. Assume $A \mid \UNI \leq A \land
\UNI$ but $A \mid \UNI \not\leq A$. Then $A \land \UNI \leq A$, so $A \mid
\UNI \leq A \land\UNI \leq A$ so $A \mid\UNI \leq A$---contradiction.  Since
the other conditions are typically required of any well structured
specification theory, we conclude that if the universal specification exists
then $A\mid \UNI \leq A$ is the sufficient and necessary condition for the
parallel composition being always stronger than conjunction.

There is an interesting open question left by the above observations.  We do
know that the parallel composition, under certain conditions, is stronger than
conjunction.  However conjunction fulfills all the axioms of parallel
composition in this setting (precongruence in particular).  It would be
interesting to explain more precisely what is it that makes parallel
composition stronger than conjunction, and what is the weakest parallel
composition that is interesting.

\bigskip

\Assume{cConjQuotient} Define the quotient of conjunction $B\setminus^\land A$ to be a
greatest (with respect to $\leq$) specification $X$ such that $A \land X \leq
B$.

\Assume{cParQuotient} Define the quotient of parallel composition $B\setminus^\mid A$ to be a
greatest (with respect to $\leq$) specification $X$ such that $A \mid X \leq
B$.

For the moment ignore the problem that a unique quotient (up to the equivalence induced
by $\leq$) may not exist. We will come back to this later. 

\begin{theorem} Consider arbitrary specifications $A$ and $B$.
	If $B\setminus^\land A$ and $B\setminus^\mid A$ are both defined
	(uniquely up to equivalence induced by $\leq$) then
	$B\setminus^\land A \leq B\setminus^\mid A$.
\end{theorem}

\begin{proof}
	Observe that $A \mid (B\setminus^\land A) \leq A \land
	(B\setminus^\land A) \leq B$.  So by uniqueness and maximality of
	$B\setminus^\mid A$ we must have that $B\setminus^\land A \leq
	B\setminus^\mid A$ (as the former also fulfils the condition of the latter).
\end{proof}

It is more  interesting to ask the question of quotient's existance. 

\Assume{cConjZero} Assume that conjunction has a null element, so: $A \land
\NULL \leq \NULL$.

\Assume{cDisj} Assume that $\lor$ is a total least upperbound operator in
specifications (a disjunction):

\Assume{cDisjLub} $A \leq C$ and $B \leq C$ implies that $A \lor B \leq C$

\Assume{cDisjLuba} $A \leq A \lor B$ and $B \leq A \lor B$.

\Assume{cDistribution} Assume that conjunction distributes over disjunction,
so: $(A\land X_1) \lor (A\land X_2) \leq B$ implies $A \land (X_1 \lor X_2)
\leq B$. 

\begin{theorem}
	In specification theories, where the conjunction distributes over
	disjunction (in the sense of the above) we have that quotient always
	exists for conjunction (so $B\setminus^\land A$ is uniquely defined up
	to equivalence), if disjunction (lub) is defined for
	arbitrary (also infinite) sets.
\end{theorem}

\begin{proof}
	First observe that \NULL\ is always a good quotient candidate. 

	If there is more than one quotient candidate then a least upper bound
	of all of them also fulfils the definition.  So we take the quotient to be the
	least upper bound of all those that fulfill the definition.%
\end{proof}

\noindent We have been unable to find generic conditions for existence of quotient for
parallel composition.

\bigskip

We shall now investigate associativity of conjunction and parallel
composition.  Order theory tells us that conjunction is associative:

\begin{theorem}
	$(A \land B) \land C \leq A \land (B \land C)$
\end{theorem}

\begin{proof}
	By A7: $(A \land B) \land C \leq C$

	By A7 again: $(A \land B) \land C \leq A \land B \leq A$

	Similarly: $(A \land B) \land C \leq B$

	By A8: $(A \land B) \land C \leq (B\land C)$

	By A8 again: $(A \land B) \land C \leq  A \land (B\land C)$
\end{proof}

To investigate associativity of parallel composition we additionally assume that 
parallel composition is idempotent:

\Assume{cParIdempotent} For all specifications $A\in\SPEC$ have $A \leq A\mid
A$.

\smallskip\noindent Note that the opposite $A\mid A \leq A$ already follows from Theorem~\ref{thm:main} and
idempotence of conjunction.
Now we obtain the following:

\begin{theorem}
	For all specifications $A$, $B$, and $C$ we have
	$(A \mid B) \mid C \leq A \mid (B \mid C)$.
\end{theorem}

\begin{proof}

	$B\mid C \geq (A\mid B)\mid C$ by Theorem~\ref{thm:main}, A7 and
	precongruence. 

	$A \mid (B\mid C) \geq A \mid [ (A\mid B)\mid C] \geq (A\mid B)\mid [
	(A\mid B)\mid C]  \geq [(A\mid B)\mid C] \mid [(A\mid B)\mid C] \geq
	(A\mid B)\mid C $.  The last step is by idempotence of $'\mid'$.
\end{proof}

Observe that the above theorem, precongruence and commutativity of parallel composition also
implies that $A \mid (B \mid C) \leq (A  \mid B) \mid C$, which completes the
expected property of associativity:

$$A\mid (B\mid C) \leq (C\mid B) \mid A \leq C \mid (B\mid A) \leq (A\mid B)
\mid C$$

\section{In a Two-Player Game Setting}\label{sec:two-players}

It turns out that the assumptions of the previous section do not hold for a
particular class of theories---those based on two player games. %
%
For example
there does not exist a universal interface automaton \cite{dAH01}
while a universal timed specification \cite{DBLP:conf/hybrid/DavidLLNW10}, if any was proposed, would
not fulfill $A \mid \UNI \leq A$.  Let us explain below, why this is the case.

Specifications with a game semantics, separate actions into controllable by
the component (outputs) and by the environment (inputs).  In such settings the
parallel composition is only possible for two components, for which the
controllable parts of the alphabet do not overlap---otherwise we have a
control conflict.  This is not required for conjunction, so conjunction exists
more often than parallel composition in such theories (and assumption A10 is
violated).  In particular this
means that $A \mid \UNI$ would typically need to have both an alphabet
identical to $A$ and different than $A$ (for example in interface automata).

Moreover an optimistic parallel composition \cite{dAH01}  is specified as a
maximum wining strategy for the input player (the player corresponding to the
context of the composition) in a safety game.  Dually conjunction is
constructed as a maximum winning strategy for the output player (the player
representing the conjunction component itself) in a safety game.   This
duality expresses very well the difference between the two operators.  The
former is concerned with correct use of the result---and use is the domain of
the context per se.  The latter is concerned with the realization of two
specifications---this clearly should be resolved within the component, and not
within the environment. 

We have fully developed this principle, when working on a real time
specification theory \cite{DBLP:conf/hybrid/DavidLLNW10}.

\section{Concluding Remarks and Future Work}\label{sec:conclusion}

The remarks placed in this short paper are clearly very preliminary.  It is
our intention to investigate more in depth algebraic structures underlying
specification theories, and use this study to better survey existing theories,
and to systematize the design of the new ones.

\bibliographystyle{eptcs} 
\bibliography{references}
\end{document}